\documentclass[aps,
showpacs,twocolumn,groupedaddress]{revtex4}

\usepackage{amsmath,amsfonts,amssymb,mathrsfs}

\begin{document}
\bibliographystyle{apsrev}


\vspace*{-1.85cm}
\begin{flushright}
hep-ph/0109264\\ TUM-HEP-438/01
\end{flushright}

\title{See-saw Mechanisms for Dirac and Majorana Neutrino Masses}

\author{Manfred Lindner}
\email[E-mail address: ]{lindner@ph.tum.de}
\affiliation{Institut f{\"u}r Theoretische Physik, Physik-Department,
Technische Universit{\"a}t M{\"u}nchen (TUM), James-Franck-Stra\ss{}e,
85748 Garching bei M{\"u}nchen, Germany}

\author{Tommy Ohlsson}
\email[E-mail address: ]{tohlsson@ph.tum.de}
\affiliation{Institut f{\"u}r Theoretische Physik, Physik-Department,
Technische Universit{\"a}t M{\"u}nchen (TUM), James-Franck-Stra\ss{}e,
85748 Garching bei M{\"u}nchen, Germany}

\author{Gerhart Seidl}
\email[E-mail address: ]{gseidl@ph.tum.de}
\affiliation{Institut f{\"u}r Theoretische Physik, Physik-Department,
Technische Universit{\"a}t M{\"u}nchen (TUM), James-Franck-Stra\ss{}e,
85748 Garching bei M{\"u}nchen, Germany}

\date{\today}

\begin{abstract}
We investigate the see-saw mechanism for generally non-fine-tuned $n
\times n$ mass matrices involving both Dirac and Majorana neutrinos. We
specifically show that the number of naturally light neutrinos cannot
exceed half of the dimension of the considered mass matrix. Furthermore,
we determine a criterion for mass matrix textures leading to light Dirac
neutrinos with the see-saw mechanism. Especially, we study $4 \times
4$ and $6 \times 6$ mass matrix textures and give some examples in
order to highlight these types of textures. Next, we present a
model scheme based on non-Abelian and discrete symmetries fulfilling the
above mentioned criterion for light Dirac neutrinos. Finally, we
investigate the connection between symmetries and the invariants of a
mass matrix on a formal level.
\end{abstract}

\maketitle

\section{Introduction}
\label{sec:intro}

Neutrino mass squared differences have lately been more and more
accurately measured by neutrino oscillation experiments. The latest
values can be summarized as follows
\cite{take00,tosh00,Toshito:2001dk,Smy:2001wf}
\begin{eqnarray}
&& \Delta m^2_{\odot} \in (10^{-5},10^{-4}) \, {\rm eV}^2, \nonumber\\
&& \Delta m^2_{\rm atm.} \simeq 2.5 \cdot 10^{-3} \, {\rm eV}^2, \nonumber
\end{eqnarray}
where $\Delta m^2_{\odot}$ is the solar mass squared difference of the
preferred large mixing angle (LMA) solution of the solar neutrino
problem and $\Delta m^2_{\rm atm.}$ is the atmospheric mass squared
difference. These results were originally obtained in two flavor
neutrino oscillation analyses, and are approximately valid in three
flavor neutrino oscillation models at least as long as the vacuum
mixing angle $\theta_2 \equiv \theta_{13}$ is small.
This means that a three flavor neutrino oscillation model
decouples into two two flavor neutrino scenarios. An upper bound of the
vacuum mixing angle $\theta_2$ has been found by the CHOOZ experiment
\cite{apol99}, $\sin^2 2 \theta_2 \lesssim 0.10$, indicating that it
is indeed small.

Neutrino oscillations depend only on the mass squared differences and
the absolute neutrino mass scale is only bound from above to about
$3 \, {\rm eV}$
\cite{groo00,Weinheimer:1999tn,Bonn:2001tw,Lobashev:1999tp,Lobashev:2001uu}.
It is also unknown if neutrinos are Dirac or 
Majorana particles \cite{bile80,doi81}. Neutrino mass models, on
the other side, depend crucially on the absolute neutrino mass scale
and on the question whether the neutrinos are Majorana or Dirac 
particles. Small Majorana neutrino masses are, for example, naturally
understood by the canonical see-saw mechanism \cite{gell79,yana79,moha80}, 
involving right-handed neutrinos with a Majorana mass matrix $M_R$
with entries which are much heavier than the electroweak scale $\epsilon$.
After integrating out the superheavy right-handed neutrinos, the 
{\it effective neutrino mass matrix} $M_\nu$ is given in terms of the 
Dirac mass matrix $M_D$ and $M_R$ as $M_\nu = - M_D M_R^{-1}M_D^T$. 
The Majorana mass matrix $M_R$ is, however, in general unrelated to 
the Dirac mass matrix $M_D$, resulting in lack of predictivity 
\cite{barr000}. This is different for models where certain symmetries 
enforce specific correlated textures for both $M_D$ and $M_R$. 
This has, for example, been achieved by introducing a conserved
U(1) charge, {\it{e.g.}}, the lepton number $\tilde{L} = L_e - L_\tau$ 
in a minimal left-right symmetric model \cite{eck87,bran88}. In general,
Abelian horizontal U(1) symmetries have been widely used in
string-inspired models \cite{iba93} of Froggatt--Nielsen type
\cite{frog79} for hierarchical neutrino masses \cite{drei95,leon96} in
order to accommodate the observed large $\nu_{\mu}$-$\nu_{\tau}$ mixing
\cite{lola99}. Though Abelian flavor symmetries tend to exhibit mixings
staying maximal under renormalization group running \cite{ell99},
maximal and bimaximal mixings appear more generically in models with 
non-Abelian flavor symmetries \cite{pak78,wilc79,gel84}. 
A drawback of non-Abelian symmetries is that the resulting neutrino 
mass matrices have typically entries of equal magnitude \cite{barr000}, 
which tends to result in degenerate neutrino masses \cite{tan99}. 
Hence, a phenomenologically successful scenario requires that these 
degeneracies are broken to a certain extent. An interesting way how
this could be achieved is if two degenerate neutrinos combine to one
quasi-Dirac neutrino. Thus, it is possible that see-saw mechanism schemes for 
Dirac neutrinos based on discrete or non-Abelian symmetries provide 
a natural link between (bi-) maximal mixing and hierarchical neutrino 
masses.

In this paper, we will investigate the types of neutrino mass matrix 
textures allowing a see-saw mechanism for both Dirac and Majorana
neutrinos when fine-tuning is absent. Different from earlier
approaches, we will not assume some conserved U(1) charge from the
beginning, which is assigned in flavor basis
\cite{eck87,bran88}. Furthermore, we will not assume additional 
hierarchies between the entries of the Dirac mass matrix $M_D$ or the
Majorana mass matrix $M_R$ \cite{dutt95}, as they could, {\it e.g.},
arise from a soft breaking of lepton numbers and permutation
symmetries \cite{Grimus:2001ex,Mohapatra:2001ns}. Especially, we will
not examine the singular see-saw mechanism for generating light
sterile neutrinos
\cite{Fukugita:1991ae,Chun:1998qw,Chikira:1998qf,McKellar:2001hi,Czakon:2001we}
in connection with Dirac neutrinos.
Instead, we will consider the most general description of textures
yielding small neutrino masses solely provided by the see-saw
mechanism (see-saw suppressed eigenvalues) and we will then discuss
the connections to symmetries arising from such see-saw suppressions.

The paper is organized as follows:
In Sec.~\ref{sec:seesaw}, we discuss some properties 
of the see-saw mechanism and the resulting mass spectrum. Furthermore, 
we will discuss the relations between fine-tuning and the principal
invariants of a neutrino mass matrix. In Sec.~\ref{sec:textures}, the
Zel'dovich--Konopinski--Mahmoud (ZKM) and pseudo-Dirac neutrinos are shortly
revisited before the concept of see-saw-Dirac particles ({\it e.g.}
neutrinos) is introduced. Then, it is shown that a $3 \times 3$ mass
matrix cannot describe a see-saw-Dirac particle, {\it i.e.}, it cannot
provide a see-saw mechanism for a Dirac particle. Next, $4 \times 4$ 
and $6 \times 6$ neutrino mass matrix textures for see-saw-Dirac
particles are discussed. In the end of Sec.~\ref{sec:textures}, we 
present a model scheme for see-saw-Dirac neutrinos in the presence 
of non-Abelian and discrete symmetries as well as algebraic relations. 
In Sec.~\ref{sec:si}, we investigate the connection between symmetries
and the principal invariants of a neutrino mass matrix on a formal
mathematical level. In the end of this section, we examine this
connection for the case of $4 \times 4$ neutrino mass matrices.
Finally, in Sec.~\ref{sec:sc}, we summarize and give our conclusions.

\section{The See-Saw Mechanism}
\label{sec:seesaw}

\subsection{Naturally Small Neutrino Masses}
\label{sec:nsnm}

The most widely accepted mechanism for the generation of small neutrino masses 
is the canonical see-saw mechanism \cite{gell79,yana79,moha80}. It involves 
the only spontaneously generated mass scale of the Standard Model (SM), 
{\it i.e.}, the electroweak scale, which is of the order
$\epsilon \sim 10^2 \; {\rm GeV} - 10^3 \; {\rm GeV}$, 
and a large mass scale which is typically of the order
$\Lambda \sim 10^{10} \; {\rm GeV} - 10^{16} \; {\rm GeV}$
or even as high as the Planck scale ($\sim 10^{19} \; \text{GeV}$),
{\it i.e.}, we have the hierarchy
\begin{equation}
\label{eq:FineTuning1}
0 < \epsilon \ll \Lambda.
\end{equation}
The complex symmetric neutrino mass matrix $M$ takes in flavor basis 
$\Psi = \left( \begin{array}{cccccc} \nu_{a,1} & \ldots & \nu_{a,n_a} &
\nu_{s,1} & \ldots & \nu_{s,n_s} \end{array} \right)^T$
the following form
\begin{equation}
\label{eq:Prop1}
M = \begin{pmatrix} 0 & M_D\\ M_D^T & M_R \end{pmatrix},
\end{equation}
where $n_a$ denotes the number of active neutrinos (in the SM, $n_a =
3$), which are elements of ${\rm SU}(2)_L$ doublets, and $n_s$ denotes
the number of sterile (singlet) neutrinos. Thus, $M$ is an $n \times
n$ matrix with $n = n_a + n_s$.
Furthermore, in Eq.~(\ref{eq:Prop1}), ``0'' denotes the $n_a \times n_a$ null 
matrix. The elements of the $n_a \times n_s$ Dirac mass matrix $M_D$ 
arise from electroweak symmetry breaking and are thus
of order $\epsilon$. The elements of the ``heavy'' $n_s \times n_s$ 
Majorana mass matrix $M_R$ are not forbidden by symmetry. These elements 
are therefore typically of order $\Lambda$, a scale provided by a
grand unified theory (GUT) or 
some other embedding which is associated with the breaking of $B - L$
symmetry.

For $n_a = n_s = 1$, {\it i.e.}, $n = 2$, the diagonalization of the neutrino 
mass matrix in Eq.~(\ref{eq:Prop1}) yields a superlight Majorana neutrino 
with mass of order $\epsilon^2/\Lambda$ and a superheavy Majorana neutrino
with mass of order $\Lambda$. The smallness of the neutrino masses follows 
from the hierarchy in Eq.~(\ref{eq:FineTuning1}), which does not
constitute a fine-tuning of the model parameters, since the presence
of the large mass scale $\Lambda$ is expected on grounds of GUTs. This
is the famous see-saw mechanism \cite{gell79,yana79,moha80} in its
simplest form, which can be generalized to $n > 2$. Note, however,
that the results will depend crucially on the specific form of the Dirac
and Majorana mass matrices $M_D$ and $M_R$. Both these matrices are
expected to emerge from scenarios involving flavor symmetries and
their breakings, which lead, for example, to so-called ``texture
zeros''. A non-trivial flavor structure can have profound consequences
and it is in general {\it not} true that the super-light neutrinos
arising from the see-saw mechanism must be Majorana neutrinos.
Instead, appropriate symmetries imposed on the fermions (and the Higgs 
fields) can, for example, enforce a texture of the mass matrix in
Eq.~(\ref{eq:Prop1}), which allows the combination of two superlight
Majorana neutrinos with opposite signs of the mass eigenvalues into
one superlight Dirac neutrino.

\subsection{Perturbation Theory and the Number of Small Neutrino Masses}

Diagonalization of the complex symmetric mass matrix $M$ given in 
Eq.~(\ref{eq:Prop1}) yields the block-diagonal form
\begin{eqnarray}
&& {\cal M} = U^T M U = \left( \begin{array}{cc} M_1 & 0 \\ 0
& M_2 \end{array} \right), \label{eq:ssm_d}
\end{eqnarray}
where $U$ is a unitary $n \times n$ matrix and $M_1$ and $M_2$ are 
$n_a \times n_a$ and $n_s \times n_s$ matrices, respectively.
The hierarchy in Eq.~(\ref{eq:FineTuning1}) allows us to consider 
the Dirac mass matrix $M_D$ in the neutrino mass matrix $M$ in 
Eq.~(\ref{eq:Prop1}) as a small perturbation of the ``unperturbed'' 
matrix, where the Majorana mass matrix $M_R$ is kept and $M_D = 0$.
Therefore, we will choose for the unitary matrix $U$ as an ansatz
\begin{equation}
U = \left( \begin{array}{cc} C_1 & S_2^\dagger \\ -S_1 & C_2^\dagger
\end{array} \right),
\end{equation}
where $C_1$ is an $n_a \times n_a$ matrix, $C_2$ is an $n_s \times
n_s$ matrix, $S_1$ and $S_2$ are $n_s \times n_a$ matrices, and the
entries of the matrices $S_i$ ($i = 1,2$) are much smaller than those of the
matrices $C_i$ ($i = 1,2$) \footnote{The entries of the matrices $S_i$
are of the order $\epsilon/\Lambda$, whereas the entries of the
matrices $C_i$ are of the order $1$.}.
Using the unitarity condition for the matrix $U$, $U^\dagger U
= U U^\dagger = 1_n$, we find that the matrices $C_i$ and $S_i$ 
have to obey $C_1^\dagger C_1 + S_1^\dagger S_1 = 1_{n_a}$, $C_2
C_2^\dagger + S_2 S_2^\dagger = 1_{n_s}$, $C_1 C_1^\dagger + S_2^\dagger
S_2 = 1_{n_a}$, $S_1 S_1^\dagger + C_2^\dagger C_2 = 1_{n_s}$, $S_2 C_1 - C_2
S_1 = 0$, and $C_1 S_1^\dagger - S_2^\dagger C_2 = 0$. Neglecting
terms that are quadratic in the matrices $S_i$ and do not appear in
combination with 
the Majorana mass matrix $M_R$, we obtain from Eq.~(\ref{eq:ssm_d})
\begin{eqnarray}
M_1 &=& - (C_1^T M_D S_1 + S_1^T M_D^T C_1) + S_1^T M_R S_1,
\label{eq:ssm_M1}\\
M_2 &=& C_2^\ast M_R C_2^\dagger + (S_2^\ast M_D C_2^\dagger + C_2^\ast
M_D^T S_2^\dagger), \label{eq:ssm_M2}\\
S_1 &\simeq &M_R^{-1} M_D^T C_1.\label{eq:ssm_SC}
\end{eqnarray}
Note that we also obtain $S_1 \simeq (M_R^{-1})^T M_D^T C_1$ which
together with Eq.~(\ref{eq:ssm_SC}) means that $M_R^T = M_R$.
Furthermore, using Eq.~(\ref{eq:ssm_SC}) and the relation $S_2 C_1 -
C_2 S_1 = 0$, we find that
\begin{equation}
S_2 \simeq C_2 M_R^{-1} M_D^T.
\label{eq:ssm_SC2}
\end{equation}
Inserting Eqs.~(\ref{eq:ssm_SC}) and (\ref{eq:ssm_SC2}) into
Eqs.~(\ref{eq:ssm_M1}) and (\ref{eq:ssm_M2}) and also using the fact
that the Majorana mass matrix $M_R$ is symmetric gives
\begin{eqnarray}
M_1 &\simeq& - C_1^T M_D M_R^{-1} M_D^T C_1, \\
M_2 &\simeq& C_2^\ast M_R C_2^\dagger \nonumber\\
&+& C_2^\ast \left( (M_R^{-1})^\ast
M_D^\dagger M_D + M_D^T M_D^\ast (M_R^{-1})^\ast \right)
C_2^\dagger. \nonumber\\
\end{eqnarray}
Since the entries of the matrix $M_D$ are much smaller than those of
the matrix $M_R$, which is consistent with Eqs.~(\ref{eq:ssm_SC}) and
(\ref{eq:ssm_SC2}), we find that
\begin{eqnarray}
M_1 &\simeq& - C_1^T M_D M_R^{-1} M_D^T C_1, \\
M_2 &\simeq& C_2^\ast M_R C_2^\dagger.
\end{eqnarray}
In the limit $M_D \to 0$, we can choose $C_1 = 1_{n_a}$ and $C_2 =
1_{n_s}$, {\it i.e.}, after block-diagonalization
the mass matrices can to lowest order in the inverse see-saw scale
$\Lambda^{-1}$ be written as
\begin{eqnarray}
M_1 &\simeq& - M_D M_R^{-1} M_D^T ,\label{eq:ssm_Mnu} \\
M_2 &\simeq& M_R. \label{eq:ssm_MnuR}
\end{eqnarray}
The matrix $M_\nu \equiv - M_D M_R^{-1} M_D^T$ on the 
right-hand side of Eq.~(\ref{eq:ssm_Mnu}) 
is an {\it effective mass matrix} 
obtained from integrating out the heavy degrees of freedom 
represented by the heavy Majorana mass matrix $M_2 \simeq M_R$. 
However, the fact that the elements of the matrices $M_D$ and $M_R$ are of the
orders $\epsilon$ and $\Lambda$, respectively, together with 
Eq.~(\ref{eq:ssm_Mnu}) does not imply $n_a$ ``see-saw mass eigenvalues'' 
of superlight Majorana neutrinos with masses of order $\epsilon^2/\Lambda$.
Similarly, Eq.~(\ref{eq:ssm_MnuR}) does not imply $n_s$ mass eigenvalues
of order $\Lambda$ for superheavy Majorana neutrinos.
The diagonalization of the $n \times n$ mass matrix in
Eq.~(\ref{eq:Prop1}) leads instead to the following pattern of 
eigenvalues.
First, for a given Majorana mass matrix $M_R$ with entries of order
$\Lambda$ there are $r \equiv \text{rank}(M_R) \leq n_s$ eigenvalues
of order $\Lambda$. Then, block diagonalization of the $n_s\times
n_s$ submatrix $M_R$ leads to an $r$ dimensional block of rank $r$
with eigenvalues of order $\Lambda$, which is placed in the down-right
corner of an $n_s\times n_s$ null matrix, whereas $n_s - r$ dimensions
of $M_R$ are not of order $\Lambda$. This can be used to divide the
complete mass matrix $M$ into blocks according to the magnitude of the
entries: First, there is the $r$ dimensional (diagonal) block of order
$\Lambda$. Then, there is the complementary diagonal block with
dimension $n_a + n_s - r$ and the off-diagonal blocks, all with
elements which are maximally of order $\epsilon$. The $n_a + n_s - r$
dimensional light block on the main diagonal is composed of the $n_a$
dimensional null matrix of the original matrix in Eq.~(\ref{eq:Prop1})
and elements of order $\epsilon$, arising from the re-organization
into the light and heavy sectors.
Thus, unless there exist specific structures in the mass matrix $M$,
which lead to cancellations due to symmetries, the $n_a+n_s-r$
dimensional block on the main diagonal naturally yields $2(n_s-r)$ or
$2n_a$ mass eigenvalues of order $\epsilon$, depending on the sign of
$n_a-n_s+r$. Written in a more compact form, there are in total
$$
e \equiv n - r - |n_a - n_s + r|
$$
mass eigenvalues of order $\epsilon$ in the $n_a + n_s - r$
dimensional light diagonal block. Including the remaining off-diagonal
blocks with elements of order $\epsilon$ does not change this result,
which can, for example, be seen by treating these blocks as
perturbations to the stiff diagonal blocks. The remaining
$$
z \equiv n - r - e = |n_a - n_s + r|
$$
mass eigenvalues are not of orders $\epsilon$ or $\Lambda$, {\it
i.e.}, they are see-saw mass eigenvalues of order
$\epsilon^2/\Lambda$, exact zeros, or further suppressed eigenvalues
of order $\epsilon^{k+1}/\Lambda^k$, where $k > 1$. With this we
arrive at the important result: {\it The number of small mass
eigenvalues naturally generated by the see-saw mechanism cannot exceed
$|n_a - n_s + r|$}. For $n_a = n_s$ this implies, for example, that
the number of see-saw mass eigenvalues is always equal to or smaller
than half of the dimension of the mass matrix $M$. This means, for
example, that it is impossible to obtain four or five see-saw mass
eigenvalues of the order $\epsilon^2/\Lambda \ll \epsilon$ from a $6
\times 6$ mass matrix $M$. Note, however, that the presence of
symmetries may further reduce the order of magnitude of the
eigenvalues, which will be discussed below.

\subsection{Fine-Tuning, Principal Invariants, and Generic Mass Scales}
\label{sec:finetuning}

We have so far discussed the natural eigenvalue spectrum of
a mass matrix with the structure in Eq.~(\ref{eq:Prop1})
without specifying any structural details of the $M_D$ and
$M_R$ matrices, which can arise from flavor symmetries and 
their breakings. Such symmetries are expected to exist and 
they lead, for example, to so-called ``texture zeros'' or other exact 
algebraic relations between different matrix elements. It is 
important to observe that flavor symmetries can (but need 
not) change the discussed generic mass eigenvalue spectrum such 
that one or more of the eigenvalues do not assume their natural 
order of magnitude. This means that an eigenvalue may turn out, 
for example, to be of order $\epsilon$ instead of order $\Lambda$, 
of order $\epsilon^2/\Lambda$ instead of order $\epsilon$, or 
$0$ instead of order $\epsilon^2/\Lambda$. Since 
$\epsilon\ll \Lambda$, this leads to a drastic change in the
order of magnitude of the corresponding eigenvalue. An 
eigenvalue which is many orders of magnitude smaller than its 
natural order of magnitude may thus be understood in terms of some
symmetry in the given mass matrix. Without such a symmetry, such a 
drastic deviation from the natural order of magnitude in the mass
eigenvalue spectrum requires a fine-tuning of parameters. 
This relation between deviations from the natural mass eigenvalue
spectrum and flavor symmetries will be further discussed in Sec.~\ref{sec:si}.

Other (in some sense also more natural) quantities for the discussion of 
the properties of the mass matrices are their invariants. The 
basis independent principal invariants $T_i$ of the mass matrix 
$M$, where $i=1,2,\dots, n$, are defined by the characteristic equation
\begin{equation}
\det (M-\lambda 1_{n}) = \lambda^{n}+\sum_{i=1}^{n}T_i\lambda^{n-i}
\end{equation}
and can be entirely determined from the mass eigenvalues alone.
In the same way as the block structure of the neutrino mass matrix 
$M$ given in Eq.~(\ref{eq:Prop1}) naturally leads to a generic 
neutrino mass eigenvalue spectrum, each of the invariants is 
characterized by generic powers of the mass scales $\epsilon$ and 
$\Lambda$. Without the block structure in Eq.~(\ref{eq:Prop1}), 
which is related to the representations of the fields under
${\rm SU}(2)_L \times {\rm U}(1)_Y$, one would expect all
entries of the matrix $M$ to be of order $\Lambda$ and the natural
scale of all invariants would therefore be $T_i =
{\cal O}(\Lambda^i)$. The presence of gauge symmetries imposes,
however, the block structure in Eq.~(\ref{eq:Prop1}) and we have a
first example where symmetries change the natural order of magnitude
of the invariants $T_i$.

The discussion of the see-saw mechanism above provided the 
natural neutrino mass scales of the orders $\epsilon^2/\Lambda$,
$\epsilon$, and $\Lambda$. We can now discuss, in a similar way,
the generic mass scales of the invariants $T_i$. 
If we denote by $T_{i,r}$ the invariant $T_i$ of a mass matrix
with $r$ mass eigenvalues of order $\Lambda$ and further 
elements of the mass matrix of order $\epsilon$, then it is easy 
to see that one obtains for the generic scales
\begin{eqnarray}
T_{i,r} &=& \epsilon^{i-r}\Lambda^r \quad \text{for} \quad r < i,\\
T_{i,r} &=& \Lambda^i \quad \text{for} \quad r \geq i.
\end{eqnarray}
For large $n_a$ and $n_s$ the specific structure of the mass matrix $M$
in Eq.~(\ref{eq:Prop1}) reduces, however, for $r < i$ the power of
$\Lambda$ by one unit (balanced by an extra factor of $\epsilon$)
whenever one of $i$ or $r$ are even, while the other one is odd. In
Table~\ref{tab:1}, the resulting generic orders of the $T_{i,r}$'s are
given for large $n_a$ and $n_s$.
\begin{table}
\begin{ruledtabular}
\begin{tabular}{c|ccccccc}
Principal & \multicolumn{7}{c}{Rank $r$} \\
invariant & 0 & 1 & 2 & 3 & 4 & 5 & 6 \\
\hline
$T_1$ & 0 & $\Lambda$ & $\Lambda$ & $\Lambda$ & $\Lambda$ & $\Lambda$
& $\Lambda$ \\
$T_2$ & $\epsilon^2$ & $\epsilon^2$ & $\Lambda^2$ & $\Lambda^2$ &
$\Lambda^2$ & $\Lambda^2$ & $\Lambda^2$ \\
$T_3$ & $\epsilon^3$ & $\epsilon^2 \Lambda$ 
& $\epsilon^2 \Lambda$ & $\Lambda^3$ & $\Lambda^3$ &
$\Lambda^3$ & $\Lambda^3$ \\
$T_4$ & $\epsilon^4$ & $\epsilon^4$ & $\epsilon^2 \Lambda^2$ 
& $\epsilon^2 \Lambda^2$ & $\Lambda^4$ & $\Lambda^4$ & $\Lambda^4$ \\
$T_5$ & $\epsilon^5$ & $\epsilon^4 \Lambda$ & 
$\epsilon^4 \Lambda$ & $\epsilon^2 \Lambda^3$ & 
$\epsilon^2 \Lambda^3$ & $\Lambda^5$ & $\Lambda^5$ \\
$T_6$ & $\epsilon^6$ & $\epsilon^6$ & $\epsilon^4 \Lambda^2$ 
& $\epsilon^4 \Lambda^2$ & $\epsilon^2 \Lambda^4$ & 
$\epsilon^2 \Lambda^4$ & $\Lambda^6$ \\
\end{tabular}
\end{ruledtabular}
\caption{The generic order of magnitude of the principal invariants
$T_i$ ($i = 1,2,\ldots,6$) of a neutrino mass matrix $M$ of arbitrary
dimension and rank $r$ ($r = 0,1,\ldots,6$) for large $n_a$ and $n_s$.}
\label{tab:1}
\end{table}
For small $n_a$ and $n_s$ these generic powers are even further
reduced. For a $4 \times 4$ mass matrix $M$ with the structure like in
Eq.~(\ref{eq:Prop1}) and $r = 2$ one obtains
\begin{eqnarray}
&& T_{1,2} = {\cal O}(\Lambda), \quad T_{2,2} = {\cal O}(\Lambda^2),
\nonumber\\
&& T_{3,2} = {\cal O}(\epsilon^2\Lambda), \quad T_{4,2} = {\cal
O}(\epsilon^4). \nonumber
\end{eqnarray}
Note that $T_{1,2}$, $T_{2,2}$, and $T_{3,2}$ are as in
Table~\ref{tab:1}, whereas $T_{4,2}$ is further reduced due to the small
values of $n_a$ and $n_s$.

In analogy with the discussion of the eigenvalues above, a deviation 
of the invariants from this generic mass scale dependence is
either a fine-tuning of parameters or a consequence of a symmetry.
Note, however, that not all symmetries lead to a change in the 
generic mass scales of the invariants. The presence or absence of 
mass scales which actually contribute to the invariants of the neutrino
mass matrix $M$ nevertheless sheds light in an interesting way 
on the symmetries constraining the neutrino mass matrix texture.

\section{Textures for See-Saw-Dirac Particles}
\label{sec:textures}

\subsection{See-Saw-Dirac Particles}
\label{sec:ssdp}

Consider in the Majorana basis a real symmetric $n\times n$ neutrino
mass matrix $M$. Let us assume that it 
can be brought via some orthogonal transformation to the block-diagonal
form
\begin{equation}\label{Eq:Superlight1}
M\mapsto\tilde{M}=
\mathscr{O}^T M\mathscr{O}=
\begin{pmatrix}
A & 0 \\
0 & B
\end{pmatrix},
\end{equation}
where $A$ denotes a real symmetric $2\times 2$ matrix given by
\begin{equation}
A=
\begin{pmatrix}
0 & \alpha\\
\alpha & 0
\end{pmatrix}
\end{equation}
and $B$ refers to some real symmetric $(n-2)\times(n-2)$ matrix. The
fields which span the matrix $A$ will be denoted by
$\left(\nu_{1L}\:\nu_{2L}\right)$.
Diagonalization of the mass matrix $\tilde{M}$ then leads to the mass matrix
\begin{equation}\label{Eq:Superlight3}
m=\text{diag}\left(\alpha,-\alpha,\beta,\gamma,\dots\right),
\end{equation}
where the two degenerate Majorana neutrinos with opposite signs of the
mass eigenvalues can be combined to one Dirac neutrino. Therefore, we
will speak of the real symmetric neutrino mass matrix $M$ as
containing a {\it Dirac particle} if diagonalization finally leads to
a mass matrix of the type given in Eq.~(\ref{Eq:Superlight3}). Note
that the mass matrix $\tilde{M}$ respects a conserved U(1) symmetry
acting on the fermionic fields with $\nu_{1L}$ carrying a charge of
$+1$ and $\nu_{2L}$ carrying a charge of $-1$.

If the mass matrix $\tilde{M}$ is identical to the mass matrix $M$
formulated in flavor basis, then the neutrinos $\nu_{1L}$ and $\nu_{2L}$ could
represent two different flavor fields. In this case, 
two Majorana neutrinos of different flavors combine to one Dirac particle and
the conserved U(1) charge is called a lepton number of the ZKM type
\cite{zel52,kon53}. If, for example, the identifications
$\nu_{1L}= \nu_{\mu L}$ and $\nu_{2L}= \nu_{\tau L}$ hold, then the mass matrix
$\tilde{M}=M$ exhibits a $\text{U}(1)_{\text{ZKM}}$ symmetry characterized by
a ZKM lepton number $\hat{L}=L_{\mu}-L_{\tau}$.

If instead $\tilde{M}\neq M$, then the fields $\nu_{1L}$ and
$\nu_{2L}$ of the matrix $A$ are not flavor fields and the basis, where
the assignment of the conserved U(1) charges (the $\nu_{1L}$-number minus the
$\nu_{2L}$-number) takes place, is different from the flavor basis. In this
case, the emerging Dirac particle is called a 
{\it pseudo-Dirac particle} \cite{wolf81}. To zeroth
order in the gauge interactions a pseudo-Dirac particle cannot be
distinguished from a genuine Dirac particle characterized by a conserved
$\text{U}(1)_{\text{ZKM}}$ lepton number. However, higher order gauge
interactions induce a splitting of the Dirac particle into two
Majorana particles with {\it nearly} degenerate masses
\cite{wolf81}. Since we will only be concerned with
zeroth order mass matrices, we will, if nothing else is mentioned, refer
to both the ZKM neutrino and the pseudo-Dirac neutrino shortly as
Dirac neutrinos.
A Dirac neutrino will be called a {\it see-saw-Dirac particle} if its mass is
superlight due to the see-saw mechanism without invoking any fine-tuning in the
sense of Sec.~\ref{sec:finetuning}.

As a quick application of the results given above, let us 
consider a $3\times 3$ mass matrix $M$, which is assumed to describe a Dirac
neutrino. If fine-tuning is absent, then this Dirac neutrino cannot have a
small mass suppressed by a see-saw mechanism, since the
necessary number of two light mass eigenvalues would exceed half of
the dimension of the mass matrix $M$, which is forbidden by the results
of Sec.~\ref{sec:finetuning}.

In the next section, we will show that see-saw-Dirac
neutrinos in the absence of any fine-tuning are first possible in the
case of $4\times 4$ mass matrices.

\subsection{$4 \times 4$ Textures}
\label{sec:4x4}

\subsubsection{Criterion for See-Saw-Dirac Particles}

According to Sec.~\ref{sec:seesaw}, we can write any real
symmetric $4\times 4$ mass matrix $M$, providing an effective see-saw
mechanism on block form like in Eq.~(\ref{eq:Prop1}), where the Dirac
mass matrix $M_D$ is given by the real $2\times 2$ matrix
\begin{equation}
 M_D\equiv
 \begin{pmatrix}
  D_1 & D_2 \\
  D_3 & D_4 
\end{pmatrix}
\end{equation} 
with $D_i = {\cal O}(\epsilon)$ or $D_i \equiv 0$, where $i=1,2,3,4$,
and the heavy Majorana mass matrix $M_R$ is given by the real symmetric 
$2\times 2$ matrix
\begin{equation}
 M_R\equiv
 \begin{pmatrix}
 R_1 & R_2 \\
 R_2 & R_3 
\end{pmatrix}
\end{equation}
with $R_i = {\cal O}(\Lambda)$ or $R_i \equiv 0$, where $i=1,2,3$.
It is assumed that the two submatrices fulfill
$\det M_D\neq 0$ and $\det M_R\neq 0$, {\it i.e.}, $r = 2$. 

If we assume that the mass matrix $M$ describes a see-saw-Dirac
particle in the sense of Sec.~\ref{sec:ssdp}, then the real symmetric
mass matrix $M$ can be brought by some orthogonal transformation $M
\mapsto m = {\mathscr O}^T M {\mathscr O}$ to the form
\begin{equation}\label{Eq:VierMalVier1}
m=\text{diag}\left(\alpha,-\alpha,\beta,\gamma\right).
\end{equation}
The invariants $T_1$ and $T_3$ defined by the characteristic equation
\begin{eqnarray}
\det (M-\lambda 1_4) &=& \det (m-\lambda
1_4) \nonumber\\
&=& \lambda^4+T_1\lambda^3+T_2\lambda^2+T_3\lambda+T_4=0 \nonumber\\
\end{eqnarray}
are as functions of the matrix elements of the matrix $M$ given by
\begin{subequations}\label{Eq:VierMalVier2}
\begin{eqnarray}
T_1(D_1,D_2,\dots,R_3)  & = & -R_1-R_3,\label{Eq:VierMalVier2a}\\
T_3(D_1,D_2,\dots,R_3) & = &
D_1^2R_3+D_2^2R_1+D_3^2R_3+D_4^2R_1 \nonumber\\
&-&2\left(D_1D_2+D_3D_4\right)R_2,
                 \label{Eq:VierMalVier2c}
\end{eqnarray}
\end{subequations}
and as functions of the matrix elements of the diagonal matrix $m$ by
\begin{subequations}
 \begin{eqnarray}
 T_1(\alpha,\beta,\gamma)  &=&  -\beta-\gamma, \label{Eq:VierMalVier3ca}\\
 T_3(\alpha,\beta,\gamma)  &=&
 \alpha^2(\beta+\gamma). \label{Eq:VierMalVier3cb}
 \end{eqnarray}
\end{subequations}
Equation~(\ref{Eq:VierMalVier3ca}) and (\ref{Eq:VierMalVier3cb}) imply that
\begin{equation}\label{Eq:VierMalVier4}
T_3=-\alpha^2 T_1. 
\end{equation} 
Assume that $T_1\neq 0$. Then, it follows from
Eq.~(\ref{Eq:VierMalVier4}) that 
\begin{equation}\label{Eq:SeesawRelation2}
 \alpha^2= - \frac{T_3(D_1,D_2,\dots ,R_3)}{T_1(D_1,D_2,\dots
 ,R_3)}\ll\epsilon^2
\end{equation}
and from Eqs.~(\ref{Eq:VierMalVier2a}) and (\ref{Eq:VierMalVier2c}) we find
that the numerator is of order $\epsilon^2 \Lambda$ and the
denominator is of order $\Lambda$. Hence, it is clear that the
relation~(\ref{Eq:SeesawRelation2}) can only be fulfilled in the presence of
some fine-tuned cancellations between the matrix elements of $M$. If we
reject fine-tuning, then it follows that $T_1=0$, {\it i.e.},
{\it the mass matrix must be traceless}. The tracelessness of the mass matrix
implies that $m=\text{diag}(\alpha,-\alpha,\beta,-\beta)$, {\it i.e.},
the four Majorana neutrinos combine to two Dirac neutrinos: One see-saw-Dirac
neutrino and one heavy Dirac neutrino. Thus,
Eq.~(\ref{Eq:VierMalVier4}) reads
\begin{equation}\label{Eq:VierMalVier6}
T_1 = T_3=0.
\end{equation}

The above considerations tell us that Eq.~(\ref{Eq:VierMalVier6}) together with
the auxiliary relations $\det M_D\neq 0$ and 
$\det M_R\neq 0$ represent necessary
and sufficient conditions for superlight Dirac neutrinos in the absence
of fine-tuning. We will therefore refer to Eq.~(\ref{Eq:VierMalVier6}) as the
{\it criterion for see-saw-Dirac particles}. Since the mass matrix $M=(M_{ij})$
is originally defined in flavor basis, there is a fundamental
difference between matrices whose matrix elements can be considered as
independent parameters and those matrices where the matrix elements are
related by some specific exact algebraic relations.

\subsubsection{Textures with Independent Entries}
\label{M4}

Under the assumption that the matrix elements $M_{ij}$ are either
exact texture zeros or independent parameters, we will now determine all
possible textures of the mass matrix $M$ in Eq.~(\ref{eq:Prop1}), which
describe a see-saw-Dirac particle. Since we will treat the matrix
elements, which are not texture zeros, as independent parameters, the
tracelessness of the mass matrix $M$ can only be fulfilled if
$R_1=R_3=0$. Thus, we obtain from Eq.~(\ref{Eq:VierMalVier2c}) that
the criterion for see-saw-Dirac particles~(\ref{Eq:VierMalVier6}) reduces to
\begin{equation}\label{Eq:AbleitungVierMalVier2}
 D_1D_2+D_3D_4 = 0.
\end{equation}
Since the matrix elements can be varied independently, the condition
(\ref{Eq:AbleitungVierMalVier2}) together with $\det M_D\neq 0$
can only be fulfilled if $D_1=D_4=0$ or $D_2=D_3=0$. 
Thus, we arrive at the conclusion that up to trivial permutations all
$4\times 4$ textures with independent non-vanishing entries, describing a
superlight Dirac particle, are all of the canonical form
\begin{equation}\label{Eq:4Mal4Kan2}
M\equiv
\begin{pmatrix}
0 & 0 & 0 & \epsilon_1\\
0 & 0 & \epsilon_2 & 0\\
0 & \epsilon_2 & 0 & \Lambda\\
\epsilon_1 & 0 & \Lambda & 0
\end{pmatrix}
\end{equation}
with $\epsilon_1,\epsilon_2={\cal O}(\epsilon)$.
Textures equivalent to the one displayed in Eq.~(\ref{Eq:4Mal4Kan2}) have been
obtained within left-right-symmetric and horizontal models implementing a
conserved horizontal U(1) charge of the ZKM type \cite{shan85}.

\subsubsection{Textures in Presence of Algebraic Relations}
\label{4Mal4Algebra}

We will now investigate the $4 \times 4$ textures leading to see-saw-Dirac
particles if algebraic relations between the matrix elements $M_{ij}$ are
allowed. If we assume $R_1=R_3=0$, but allow for algebraic relations
within the Dirac matrix $M_D$, then the criterion for see-saw-Dirac particles
is still given by Eq.~(\ref{Eq:AbleitungVierMalVier2}) and the
corresponding texture is equal to
\begin{equation}
M=
\begin{pmatrix}
 0 & 0 & \epsilon_1 & \epsilon_2 \\
 0 & 0 & -\frac{\epsilon_1\epsilon_2}{\epsilon_3} & \epsilon_3\\
 \epsilon_1 & -\frac{\epsilon_1\epsilon_2}{\epsilon_3} & 0 & \Lambda \\
 \epsilon_2 & \epsilon_3 & \Lambda & 0
\end{pmatrix}
\end{equation}
with $\epsilon_i={\cal O}(\epsilon)\ll{\cal O}(\Lambda)$, where $i=1,2,3$.
Let us now assume that the relation $R_1 = - R_3 \neq 0$ holds.
Then, the criterion for see-saw-Dirac particles (\ref{Eq:VierMalVier6})
reads
\begin{eqnarray}
T_3 &=& R_1\left(-D_1^2+D_2^2-D_3^2+D_4^2\right) \nonumber\\
&-&2R_2(D_1D_2+D_3D_4) = 0.
\label{Eq:4Mal4Algebra9}
\end{eqnarray}
If we assume that $R_1$ and $R_2$ can be varied independently, then
both parentheses in Eq.~(\ref{Eq:4Mal4Algebra9}) are necessarily equal to
zero. The second parenthesis vanishes for the choice $D_3=-D_1D_2/D_4$, which
means that the first parenthesis vanishes when $D_1^2=D_4^2$. Thus, we can in
this case determine the texture to be
\begin{equation}
 M=
 \begin{pmatrix}
  0 & 0  & \epsilon_1 & \epsilon_2\\
  0 & 0  &\mp\epsilon_2 & \pm\epsilon_1\\
  \epsilon_1 & \mp\epsilon_2 &  \Lambda_1 & \Lambda_2\\
  \epsilon_2 & \pm\epsilon_1  & \Lambda_2 & -\Lambda_1
 \end{pmatrix}.
\label{Eq:4Mal4Algebra10a}
\end{equation}
with $\epsilon_{1},\epsilon_{2}={\cal O}(\epsilon)$ and
$\Lambda_{1},\Lambda_2={\cal O}(\Lambda)$.
Let the mass matrix given in Eq.~(\ref{Eq:4Mal4Algebra10a}) be spanned
by the fields $\left(\nu_{1L}\:\nu_{2L}\:N_1\:N_2\right)$, where
$\nu_{1L}$ and $\nu_{2L}$ denote two active neutrinos and $N_1$ and
$N_2$ denote two SM singlets. Then, the texture in
Eq.~(\ref{Eq:4Mal4Algebra10a}) can be naturally obtained by imposing
the discrete symmetry
\begin{equation}
{\cal{D}}: \left\{
\begin{matrix}
 \nu_{1L} \rightarrow \mp {\rm i} \nu_{2L},
 & \nu_{2L} \rightarrow \pm {\rm i} \nu_{1L},\\
 N_1 \rightarrow {\rm i} N_2, & N_2 \rightarrow - {\rm i} N_1,
\end{matrix}\right.
\end{equation} 
in the Majorana basis.

\subsection{$6\times 6$ Textures}
\label{sec:6x6}

\subsubsection{Criterion for See-Saw-Dirac Particles}

In order to maximize the number of superlight particles emerging from the
see-saw mechanism, we can again according to Sec.~\ref{sec:seesaw} write
a real symmetric $6\times 6$ mass matrix on block form as in
Eq.~(\ref{eq:Prop1}), where the real Dirac mass matrix $M_D$ is given by the
$3\times 3$ matrix
\begin{equation}
 M_D\equiv
 \begin{pmatrix}
  D_1 & D_2 & D_3\\
  D_4 & D_5 & D_6\\
  D_7 & D_8 & D_9
\end{pmatrix}
\label{Eq:SechsMalSechs2}
\end{equation} 
with $D_i = {\cal O}(\epsilon)$ or $D_i \equiv 0$,
where $i=1,2,\dots,9$, and the real symmetric heavy Majorana mass matrix
$M_R$ is given by the $3\times 3$ matrix
\begin{equation}
 M_R\equiv
 \begin{pmatrix}
 R_1 & R_2 & R_3\\
 R_2 & R_4 & R_5\\
 R_3 & R_5 & R_6
\end{pmatrix}
\label{Eq:SechsMalSechs3}
\end{equation}
with $R_i = {\cal O}(\Lambda)$ or $R_i \equiv 0$, where $i=1,2,\dots,6$.
To guarantee an effective see-saw mechanism, it is additionally assumed that
$\det M_D\neq 0$ and $\det M_R\neq 0$. Let us assume that the mass matrix $M$
describes a see-saw-Dirac particle.
This means that the superlight mass eigenvalue spectrum is of the type
$\alpha,-\alpha,\beta\simeq\epsilon^2/\Lambda$, where $\alpha$ can be regarded
as the mass of the Dirac particle. Consider the characteristic
equation of the mass matrix $M$ written in block form
\begin{equation}\label{Eq:SechsMalSechs7}
 \det (M-\lambda 1_6)=
 \begin{vmatrix}
  -\lambda 1_3 & M_D\\
  M_D^{T} & M_R-\lambda 1_3 
 \end{vmatrix} 
 =0.
\end{equation}
Since $\det M_R\neq 0$, the matrix $M_R-\lambda 1_3$ can be inverted for
$\lambda\ll\Lambda$. One can therefore apply the Gauss elimination algorithm
to the block matrix $M-\lambda 1_6$, since it leaves the determinant invariant.
{}From Eq.~(\ref{Eq:SechsMalSechs7}), we therefore obtain that the superlight
mass eigenvalues of the mass matrix $M$ are also solutions of 
\begin{equation}\label{Eq:SechsMalSechs9}
 \left|-\lambda 1_3-M_D(M_R-\lambda 1_3)^{-1}M_D^{T}\right|=0.
\end{equation}
Expanding this determinant for a small parameter $\lambda$, 
Eq.~(\ref{Eq:SechsMalSechs9}) can be re-written as
\begin{eqnarray}
&& \left|-\lambda 1_3-M_D M_R^{-1} M_D^{T} + {\cal
 O}\left( \frac{\epsilon^4}{\Lambda^3} \right)\right| \nonumber\\
& = & \lambda^3 + \left[ T_1 + {\cal O}\left(
 \frac{\epsilon^4}{\Lambda^3} \right) \right] \lambda^2 + \left[ T_2 + {\cal
 O}\left( \frac{\epsilon^8}{\Lambda^6} \right) \right] \lambda
 \nonumber\\
&+& \left[ T_3 + {\cal O}\left( \frac{\epsilon^{12}}{\Lambda^9}
 \right) \right] = 0,
\end{eqnarray}
where the invariants $T_1$, $T_2$, and $T_3$ are defined by the characteristic
equation $\det
(M_{\nu}-\lambda 1_3)=\lambda^3+T_1\lambda^2+T_2\lambda+T_3=0$ of the 
effective mass matrix $M_{\nu}=-M_DM_R^{-1}M_D^{T}$.
Using $\lambda = {\cal O}(\epsilon^2/\Lambda)$,
we find that
\begin{equation}\label{Eq:SechsMalSechs9b}
\lambda^3+T_1\lambda^2+T_2\lambda+T_3={\cal{O}}\left(\frac{\epsilon^8}
{\Lambda^5}\right).
\end{equation}
Taking the previously assumed light mass eigenvalue spectrum into
account, we obtain from Eq.~(\ref{Eq:SechsMalSechs9b}) the following
equations for the mass $\alpha$ of the see-saw-Dirac particle
\begin{equation}
  T_1\alpha^2+T_3  ={\cal{O}}\left(\frac{\epsilon^8}
{\Lambda^5}\right),\quad
  \alpha^3+T_2\alpha ={\cal{O}}\left(\frac{\epsilon^8}
  {\Lambda^5}\right).
 \end{equation}
Since $\alpha\neq 0$,
this system of equations can only have a solution if 
\begin{equation}\label{Eq:SechsMalSechs10a}
T_3-T_1 T_2={\cal{O}}\left(\frac{\epsilon^8}
  {\Lambda^5}\right).
\end{equation}
Note that the invariant $T_3$ and the product $T_1 T_2$ are both of the order
$\epsilon^6/\Lambda^3$. Hence, Eq.~(\ref{Eq:SechsMalSechs10a}) expresses a
fine-tuning of the mass matrix $M$ unless the right-hand side
vanishes. Since we reject fine-tuning, the invariants must therefore {\it
exactly} fulfill
\begin{equation}\label{Eq:SechsMalSechs10}
T_3 - T_1 T_2= 0.
\end{equation}
In fact, it is easy to show that Eq.~(\ref{Eq:SechsMalSechs10}) is
also a sufficient condition for a superlight Dirac particle. Hence, we will
call it the {\it criterion for see-saw-Dirac particles}. It has thus been
shown that the first order in the inverse see-saw scale given by the effective
mass matrix $M_{\nu}$ is already sufficient to determine
non-fine-tuned neutrino mass textures leading to see-saw-Dirac particles.

\subsubsection{Textures Not Yielding See-Saw-Dirac Neutrinos}

Consider an ansatz where the Dirac mass matrix $M_D$ is equivalent to
the matrix
\begin{equation}\label{Eq:SechsMalSechsAlgebra1}
 M_D\equiv\text{diag}(D_1,D_5,D_9),
\end{equation}
being the simplest form consistent with $\det M_D\neq 0$. For this
Dirac mass matrix $M_D$ there exist {\it exactly} two inequivalent Majorana
mass matrices, which never lead to a superlight Dirac particle, even
if arbitrary algebraic relations between the matrix elements are
allowed. These matrices are
$$
M_R  = 
\begin{pmatrix}
R_1 & R_2 & 0 \\
R_2 & 0 & R_5 \\
0 & R_5 & 0
\end{pmatrix},
\qquad
M_R' =
\begin{pmatrix}
0 & R_2 & R_3 \\
R_2 & 0 & R_5 \\
R_3 & R_5 & 0
\end{pmatrix}.
$$ 
The criterion for see-saw-Dirac particles given in 
Eq.~(\ref{Eq:SechsMalSechs10}), corresponding to the Majorana mass
matrix $M_R$, reads
\begin{equation}
(T_3 - T_1 T_2)\det M_R
=D_9^4D_5^2R_1^2R_2^2R_5^2=0,
\end{equation}
which can only be fulfilled if an additional texture zero
would be introduced in the mass matrix $M$, leading to a different class of
textures or letting $\det M$ vanish. Similar considerations hold for the
Majorana mass matrix $M_R'$.

\subsubsection{Textures with Independent Entries}

Assume now that all non-vanishing matrix elements of the mass matrix $M$ are 
independent parameters. Then, all possible textures consistent with
the criterion for see-saw-Dirac particles given in
Eq.~(\ref{Eq:SechsMalSechs10}) are equivalent to the unique form
\begin{equation}
 M=
 \begin{pmatrix}
  0   & \epsilon_1       & 0   & 0   & 0         & 0  \\
  \epsilon_1 & \Lambda_1 & 0   & 0   & 0         & 0   \\
  0   & 0         & 0   & 0   & 0         & \epsilon_2 \\
  0   & 0         & 0   & 0   & \epsilon_3       & 0   \\
  0   & 0         & 0   & \epsilon_3 & 0         & \Lambda_2\\
  0   & 0         & \epsilon_2 & 0   & \Lambda_2 & 0
\end{pmatrix}
\label{Eq:SechsMalSechs12}
\end{equation}
with $\epsilon_1,\epsilon_2,\epsilon_3={\cal{O}}(\epsilon)$ and
$\Lambda_1,\Lambda_2={\cal{O}}(\Lambda)$.
{}From Eq.~(\ref{Eq:SechsMalSechs12}) we observe that the superlight particles
obtain their masses via two decoupled mechanisms:
\begin{enumerate}
\item The Majorana particle obtains its mass $\beta$ from the ordinary
$2\times 2$ see-saw mechanism.
\item The Dirac particle obtains its mass $\alpha$ from the canonical
$4\times 4$ texture given in Eq.~(\ref{Eq:4Mal4Kan2}).
\end{enumerate}
Hence, it is shown for $6\times 6$ textures with independent
entries that the model-independent method presented here to determine all
possible textures consistent with the criterion for see-saw-Dirac particles
is equivalent to the introduction of a conserved lepton
number of the ZKM type \cite{eck87,bran88}.

\subsubsection{A Model Scheme for Textures in Presence of
Non-Abelian and Discrete Symmetries}

Non-Abelian as well as discrete symmetries between leptons mostly predict 
lepton mass matrix textures, where some of the matrix elements are equal in
magnitude, which can result in maximal or bimaximal mixing and degenerate
neutrino masses \cite{barr000,tan99}. In the case of non-Abelian symmetries,
models have been proposed, where only one mixing angle being {\it{nearly}}
maximal is generic \cite{wett99}. However, most of these models have problems
to naturally accommodate large $\nu_{\mu}$-$\nu_{\tau}$ mixing and
hierarchical mass squared differences \cite{barr000}. Especially, the
presence of a ZKM lepton number in the neutrino sector as a source of
(nearly) maximal atmospheric mixing usually implies the reverse order of the
hierarchy between $\Delta m^2_{\odot}$ and $\Delta m^2_{\rm atm.}$
\cite{grimus}.
On the other hand, a combination of neutrinos to a
pseudo-Dirac-particle, in the sense of Sec.~\ref{sec:ssdp}, does not
establish a specific mixing between the corresponding generations, and
hence, allows more freedom in accommodating mixing angles and mass
squared differences.

Contrary to the usual approach, we will therefore concentrate on the
natural generation of hierarchical mass squared differences by imposing a
conserved U(1) charge and assume that the U(1) generator is not diagonal in
flavor basis, {\it i.e.}, some of the active neutrinos combine to a
pseudo-Dirac particle instead of a genuine Dirac-particle.
The pseudo-Dirac character of some of the neutrinos
requires the presence of exact algebraic relations between the matrix elements.

Let $\nu_e$, $\nu_{\mu}$, and $\nu_{\tau}$ denote the SM neutrino flavor fields
and $N_1$, $N_2$, and $N_3$ denote the charge conjugates of three right-handed
neutrinos. To be specific, we will assume in the basis
$\left(\overline{{N_1}^c}\:N_1\:\overline{{N_2}^c}\:N_2\:\overline{{N_3}^c}
\:N_3\right)$ that the corresponding $3\times 3$ Majorana mass matrix $M_R$ in
Eq.~(\ref{eq:Prop1}) is on the form \cite{ma99,barb001}
\begin{equation}
 M_R=\Lambda
 \begin{pmatrix}
  0 & 0 & 1\\
  0 &-1&0\\
  1&0&0
 \end{pmatrix}.
\label{eq:Nmass}
\end{equation}
This kind of Majorana matrix has received interest in the
framework of SO(3) flavor symmetries, where the three SM singlets
$N_1$, $N_2$, and $N_3$ are combined to an SO(3) triplet field
\cite{barb001,car98,ma99,wett99,wu991,wu992,wu993}. Invariance of the neutrino
mass terms under the discrete $Z_2$ symmetry
\begin{equation}\label{Eq:Model1}
  {\cal{D}}:
 \left\{
  \begin{matrix}
   \nu_e\rightarrow -{\rm i}\nu_{\mu},
   & \nu_{\mu}\rightarrow {\rm i}\nu_e,
    & \nu_{\tau}\rightarrow -\nu_{\tau},\\
 N_1\rightarrow {\rm i}N_3, & N_2\rightarrow -N_2, & N_3\rightarrow
  -{\rm i}N_1
 \end{matrix}\right.
\end{equation}
enforces the $6 \times 6$ neutrino mass matrix $M$ to be on the form
\begin{equation}\label{Eq:Model2}
M=
 \begin{pmatrix}
  0 & 0 & 0 & \epsilon_1 & 0 & \epsilon_2\\
  0 & 0 & 0 & -\epsilon_2 & 0 & \epsilon_1 \\
  0 & 0 & 0 & 0 & \epsilon_3 & 0\\
  \epsilon_1 &  -\epsilon_2 & 0 & 0 & 0 & \Lambda \\
  0 &  0 & \epsilon_3 & 0 & -\Lambda & 0 \\
  \epsilon_2 &  \epsilon_1 & 0 & \Lambda & 0 & 0\\
\end{pmatrix},
\end{equation}
where $\epsilon_1,\epsilon_2$, and $\epsilon_3$ denote three different real
matrix elements of the order of the electroweak scale $\epsilon$. 
The algebraic relations between some of the entries of the mass matrix
$M$ in Eq.~(\ref{Eq:Model2}), which are established by the discrete
symmetry ${\cal{D}}$, allow the existence of a see-saw-Dirac neutrino
in the absence of a conserved ZKM charge. Due to the fact that there
is no conserved ZKM charge present, radiative corrections will induce
a small splitting of the degenerate masses and could thereby, in
principle, establish the hierarchy between $\Delta m^2_{\odot}$ and
$\Delta m^2_{\rm atm.}$.

\section{The Connection between Symmetries and Invariants}
\label{sec:si}

The smallness of the neutrino masses is well understood in terms of the 
see-saw mechanism, which is a consequence of the general block form of
the neutrino mass matrix $M$ given in Eq.~(\ref{eq:Prop1}). The
sub-blocks of $M$ leave, however, the mixing angles and mass spectrum of the
light neutrinos unspecified. Predictions for mixing angles
and mass spectra of neutrinos require further horizontal or flavor symmetries 
enforcing specific textures of the Dirac and Majorana mass matrices.
Therefore, it is important to relate general properties of the mass
matrix as discussed above to the symmetries acting on the different
neutrino flavors.

Assume that there exists a (flavor symmetry) group $G$, which is mapped into a
reducible unitary representation $D$ \footnote{Note that all representations 
of finite groups are equivalent to unitary representations.} acting on
the flavor space:
\begin{equation}
g\mapsto D(g),\quad g\in G.
\nonumber
\end{equation}
The matrix representation $D(g)$ acts on the state vectors
$\Psi=\left(\begin{array}{cccccc} \nu_{a,1} & \dots & \nu_{a,n_a} &
\nu_{s,1} & \dots & \nu_{s,n_s} \end{array}\right)^T$ (here given in
flavor basis), where $n_a$ denotes again the number of active
neutrinos, which are elements of $SU(2)_L$ doublets, and $n_s$ denotes
the number of sterile neutrinos (SM singlets):
\begin{equation}\label{eq:flavorsymmetry2}
 \Psi\mapsto D(g)\Psi,\quad g\in G.
\end{equation}
Next, consider two irreducible subrepresentations $D^{(\alpha)}(g)$ and
$D^{(\beta)}(g)$ of the representation $D(g)$ and some matrix $X$
which fulfills
\begin{equation}\label{eq:Schur1}
\left( D^{(\beta)}(g) \right)^T X D^{(\alpha)}(g) = X
\end{equation}
for all $g \in G$. Using Eq.~(\ref{eq:Schur1}) and the unitarity of
the representations, we obtain that
\begin{equation}\label{eq:Schur2}
 X^\dagger X D^{(\alpha)}(g) = D^{(\alpha)}(g) X^\dagger X. 
\end{equation}
Since the representation $ D^{(\alpha)}(g)$ is irreducible, it follows from
Schur's lemma \cite{hame89,geor99} that $X^\dagger X$ is proportional
to the unit matrix, {\it i.e.}, the matrix $X$ is a {\it unitary}
matrix times some arbitrary mass scale, which is given as a physical input
parameter. Furthermore, it follows that the matrix 
$X$ is, up to this overall factor, uniquely determined by the choice of the
matrix representations $D^{(\alpha)}(g)$ and
$D^{(\beta)}(g)$. Moreover, Schur's lemma tells us that in case that
besides Eq.~(\ref{eq:Schur1}) for some irreducible representation
$D^{(\gamma)}(g)$ there holds also a relation
\begin{equation}
\left(D^{(\beta)}(g)\right)^T X D^{(\gamma)}(g) = X
\end{equation}
for all $g\in G$, the representation $D^{(\gamma)}(g)$ is equivalent to the
representation $D^{(\alpha)}(g)$.

We will now consider the case when the symmetry group $G$ is unbroken at low
energies. The origin of the mass terms in $M_R$ of the heavy sterile neutrinos
generated at the GUT or embedding scale is in general quite different from the
origin of the Dirac mass terms in $M_D$, which emerge from Yukawa couplings and
the electroweak vacuum expectation value. It can therefore be assumed
that in flavor basis the representation $D(g)$ takes on the block-diagonal form
\begin{equation}\label{eq:flavorsymmetry0}
 D(g)=D_a(g)\oplus D_s(g),
\end{equation}
where $D_a(g)$ is an $n_a$ dimensional unitary representation acting on the
subspace of the active neutrinos only and $D_s(g)$ is an $n_s$ dimensional 
representation which acts only on the sterile neutrinos.

{}From the unitarity of the representation $D(g)$ it follows that an
appropriate change of basis allows
us to write the representations $D_a(g)$ and $D_s(g)$ in reduced form as
\begin{equation}\label{eq:flavorsymmetry1}
D_a(g)=\bigoplus_{\alpha}k^a_{\alpha}D_a^{(\alpha)}(g),\quad
D_s(g)=\bigoplus_{\alpha}k^s_{\alpha}D_s^{(\alpha)}(g),
\end{equation}
where $\alpha=1,2,\dots$ runs only over the inequivalent irreducible
representations and the integer $k^x_{\alpha}$, $x=a,s$, specifies how
often the irreducible representation $D_x^{(\alpha)}(g)$ occurs
in the reduction. This means that the basis has been
chosen such that the matrices of equivalent representations are
identical and
\begin{equation}
 k^x_\alpha D_x^{(\alpha)}(g) \equiv \underbrace{D_x^{(\alpha)}(g)
 \oplus \dots \oplus D_x^{(\alpha)}(g)}_{k^x_\alpha},\quad x=a,s.
\end{equation}
Specifically, if two representations $D^{(\alpha)}_a(g)$ and $D^{(\beta)}_s(g)$
are equivalent, they are also understood to be identical and we can
therefore choose the labeling such that $\beta\equiv\alpha$.
The dimension of the irreducible representation $D_x^{(\alpha)}(g)$
will be denoted by $d^x_\alpha$, {\it i.e.}, the representation
$D_x^{(\alpha)}(g)$ is a $d^x_\alpha\times d^x_\alpha$ matrix.

Consider now the neutrino mass matrix $M$ in Eq.~(\ref{eq:Prop1}) and
assume for simplicity that the Dirac mass matrix $M_D$ as well as the
Majorana mass matrix $M_R$ are non-singular. Due to
Eq.~(\ref{eq:flavorsymmetry0}) the unitary transformation $V$ which brings
the representation $D(g)$ to the completely reduced form in
Eq.~(\ref{eq:flavorsymmetry1}) decomposes in flavor basis into
$V=V_a\oplus V_s$, where $V_a$ and $V_s$ are unitary matrices of
dimensions $n_a$ and $n_s$, respectively. Therefore,
in the basis where the representation $D(g)$ is on the completely reduced form
in Eq.~(\ref{eq:flavorsymmetry1}), the neutrino mass matrix $M'\equiv V^TMV$
reads
\begin{equation}\label{eq:M'}
M'= \begin{pmatrix} 0 & M_D'\\ {M_D'}^T & M_R' \end{pmatrix} \equiv
\begin{pmatrix} 0 & V_a^TM_DV_s\\ V^T_sM_D^TV_a & V^T_sM_RV_s \end{pmatrix}
\end{equation}
and the effective neutrino mass matrix is simply given by 
$M'_\nu\equiv -M_D'{M_R'}^{-1}{M_D'}^T$. In this basis, we can now apply the
above stated implications of Schur's lemma in order to determine the neutrino
mass matrix $M'$, following from the invariance of the neutrino mass term
$\overline{\Psi^c}M'\psi\sim\Psi^T M'\Psi$
\footnote{Here we have omitted the gamma matrices, since they only
act on the Lorentz indices, but not on the flavor indices.}
under the transformations
\begin{equation}
 \Psi^TM'\Psi\mapsto \Psi^TD^T(g)M'D(g)\Psi,\quad g\in G.
\end{equation}
As a result, we obtain that the Majorana mass matrix
$M_R'$ is, up to trivial permutations, on block-diagonal form
\begin{equation}\label{eq:Majorana}
M_R' = \text{diag}(A_1,A_2,\dots),
\end{equation}
where each submatrix $A_\alpha$, $\alpha=1,2,\dots$, defines a bilinear form
which is invariant under the symmetry $G$ in the representation
$k^s_\alpha D_s^{(\alpha)}(g)$.

If for some $\alpha=1,2,\dots$ the representation $D_s^{(\alpha)}(g)$ is
equivalent to its complex conjugate representation ${D_s^{(\alpha)}}^\ast (g)$,
then the matrix $A_\alpha$ can be written as
\begin{subequations}\label{eq:blocks} 
\begin{equation}\label{eq:block1}
A_\alpha = A'_\alpha \otimes U_\alpha,
\end{equation}
where $A'_\alpha$ denotes some arbitrary invertible symmetric
$k^s_\alpha\times k^s_\alpha$ matrix
with undetermined entries of the generic order $\Lambda$, whereas the unitary
$d^s_\alpha\times d^s_\alpha$ matrix $U_\alpha$ is determined by the choice of
the representation $D_s^{(\alpha)}(g)$ \footnote{In fact,
the unitary matrix is in this case either symmetric or antisymmetric,
depending on the question whether $D_s^{(\alpha)}$ is an {\it integer}
or a {\it half-integer} representation.}.
If instead the representation $D_s^{(\alpha)}(g)$ is {\it complex}, {\it i.e.},
not equivalent to its complex
conjugate, then the condition $\det M_R \neq 0$ requires the complete
reduction of the representation $D_s(g)$ in Eq.~(\ref{eq:flavorsymmetry1})
also to contain the complex conjugate representation ${D_s^{(\alpha)}}^\ast(g)$
exactly $k_\alpha^s$ times. Thus, the matrix $A_\alpha$ is on the form
\begin{equation}\label{eq:block2}
A_\alpha = \begin{pmatrix} 0 & A'_\alpha\\ {A'_\alpha}^T & 0 \end{pmatrix}
\otimes U_\alpha,
\end{equation}
\end{subequations}
where ``0'' denotes the $k^s_\alpha \times k^s_\alpha$ null
matrix, $A'_\alpha$ denotes some arbitrary invertible 
$k^s_\alpha\times k^s_\alpha$
matrix with undetermined entries of the order $\Lambda$, and the unitary
$d^s_\alpha\times d^s_\alpha$ matrix $U_\alpha$ is again determined by the
choice of the representation $D_s^{(\alpha)}(g)$.

As a simple example, consider the Majorana mass matrix $M_R$ in
Eq.~(\ref{eq:Nmass}).
Before symmetry breaking, the representation $D_s(g)=D_s^{(\alpha)}(g)$,
where $\alpha\equiv 1$, is the (irreducible) 3-dimensional representation
of SO(3), which is equivalent to its complex conjugate representation.
Therefore, Eq.~(\ref{eq:block1}) applies when $\Lambda$ is the
arbitrary ``matrix'' $A'_\alpha$ and
$M_R/\Lambda$ is the uniquely determined unitary (even orthogonal)
matrix $U_\alpha$.

Similarly to the treatment of the Majorana mass matrix $M_R'$ in
Eq.~(\ref{eq:Majorana}), one also verifies that the Dirac mass matrix $M_D'$
decomposes into the unitary submatrices $U_\alpha$ known from
Eqs.~(\ref{eq:block1}) and (\ref{eq:block2}) times a mass scale of the
order $\epsilon$, in such a way that the effective neutrino mass
matrix $M_\nu'$ can be determined by consistently carrying out block
multiplications. After an appropriate re-labeling of the
representations $D_a^{(\alpha)}(g)$ and $D_s^{(\alpha)}(g)$, we obtain
for the effective neutrino mass matrix the block-diagonal form
\begin{equation}\label{eq:effectivematrix}
M'_\nu = \text{diag}(B_1,B_2,\dots),
\end{equation}
where each matrix $B_\alpha$, $\alpha=1,2,\dots$, defines a bilinear form,
which is invariant under the symmetry $G$ in the representation
$k^a_\alpha D_a^{(\alpha)}(g)$.
If for some $\alpha=1,2,\dots$ the representation $D_a^{(\alpha)}(g)$ is
equivalent to its complex conjugate representation ${D_a^{(\alpha)}}^\ast (g)$,
then the matrix $B_\alpha$ reads 
\begin{subequations}\label{eq:effective}
\begin{equation}
B_\alpha = B'_\alpha \otimes U_\alpha,
\end{equation}
where $B'_\alpha$ denotes some arbitrary invertible symmetric
$k^a_\alpha\times k^a_\alpha$ matrix
with entries of the order $\epsilon^2/\Lambda$ and the unitary
$d^a_\alpha\times d^a_\alpha$ matrix $U_\alpha$ is fixed by the representation 
$D_a^{(\alpha)}(g)$.

For a complex representation $D_a^{(\alpha)}(g)$ the invertibility of
the Dirac matrix $M_D$ requires the matrix $B_\alpha$ to be on the form
\begin{equation}
B_\alpha = \begin{pmatrix} 0 & B'_\alpha\\ {B'_\alpha}^T & 0 \end{pmatrix}
\otimes U_\alpha,
\end{equation}
\end{subequations}
where ``0'' denotes the $k^a_\alpha \times k^a_\alpha$ null
matrix, $B'_\alpha$ denotes some invertible $k^a_\alpha\times k^a_\alpha$
matrix with entries of the order $\epsilon^2/\Lambda$, and the unitary
$d^a_\alpha\times d^a_\alpha$ matrix $U_\alpha$ again depends on the choice of
the representation $D_a^{(\alpha)}(g)$.
The important thing here is to note that each matrix element of the neutrino
mass matrix $M'$ in Eq.~(\ref{eq:M'}) can serve as a parameter of one and only
one of the matrices $B_\alpha$, {\it i.e.}, for
$\alpha\neq\beta$ the matrices $B_\alpha$ and $B_\beta$ are described by
decoupled mass parameters and are therefore independent in a parametrical
sense.

Taking $B_\alpha^\dagger B_\alpha$, it is readily seen that
block-diagonalization of $B_\alpha$ yields $k^a_\alpha$ different sets
of neutrino mass eigenvalues, which are, up to relative phases,
$d^a_\alpha$-fold (or, if the representation $D_a^{(\alpha)}(g)$ is complex, 
$2d^a_\alpha$-fold) {\it degenerate}.
The neutrino masses, each of which is $d^a_\alpha$-fold ($2d^a_\alpha$-fold)
degenerate, will be denoted by $m_{\alpha l}$, where $l=1,2,\dots, k^a_\alpha$.
In this case, the $k^a_\alpha$ different neutrino masses $m_{\alpha l}$
are actually correlated, {\it i.e.}, they are in a parametrical sense
dependent.
However, it is crucial to note that this correlation is only due to the
diagonalization of the matrix $B'_\alpha$, whose entries are not constrained by
the symmetry $G$, which has already been fully taken into account when
introducing the unitary matrices $U_\alpha$ in Eqs.~(\ref{eq:blocks}) and
(\ref{eq:effective}). Except of their common mass scale, which is
$\epsilon^2/\Lambda$, the $k^a_\alpha$ different masses
exhibit no further relations, which are protected by the
symmetry $G$. Hence, we can regard them as independent in a
{\it generic} sense.

Let us now specialize to the case when the neutrino mass matrix $M$
is real. The principal invariants $T_i$ $(i=1,2,\dots,n)$ of the
neutrino mass matrix $M$ can be expanded as finite sums of powers of
the electroweak scale $\epsilon$ and the GUT or embedding scale
$\Lambda$ as follows:
\begin{equation}\label{eq:expansion}
T_i = \underset{j+k=i}{\sum_{j}\sum_{k}}
a^i_{jk} \left(\frac{\epsilon^2}{\Lambda}\right)^j \Lambda^k,
\end{equation}
where the non-negative integers $j$ and $k$ have to obey $j+k=i$. In
Eq.~(\ref{eq:expansion}), the term $a^i_{jk}\epsilon^{2j}\Lambda^{k-j}$ is the
sum over all products of mass eigenvalues, where $j$ eigenvalues are of the
order $\epsilon^2/\Lambda$ and $k$ eigenvalues are of the order $\Lambda$.
Since the symmetry $G$ implies that some of the eigenvalues are up to a sign
degenerate it can happen that some (or all) of the coefficients
$a^i_{jk}$ vanish exactly.
However, since the absolute value of the neutrino masses $m_{\alpha l}$ is
undetermined by the symmetry $G$, any non-vanishing coefficient
$a^i_{jk}\neq 0$ must generically be of order unity. In other words,
a situation where $0<|a^i_{jk}|\ll 1$ cannot be understood in terms of the
symmetry $G$ and must therefore be the result of some fine-tuning of the
model parameters.

Within the above presented framework one can now even test the validity of the
criteria for see-saw-Dirac particles given in Secs.~\ref{sec:4x4} and
\ref{sec:6x6}.
First, we note that following Eq.~(\ref{eq:expansion}), the principal
invariants $T_1$ and $T_3$ of the $4\times 4$ mass matrix $M$ in
Sec.~\ref{sec:4x4} can be written as
\begin{subequations}
\begin{eqnarray}
 T_1 & = & a^1_{01}\Lambda,\\
 T_3 & = & a^3_{12}\epsilon^2\Lambda.
\end{eqnarray}
\end{subequations}
Next, using that the generic mass scale of the see-saw-Dirac particle is of
the order $\epsilon^2/\Lambda$, Eq.~(\ref{Eq:VierMalVier4}) implies that
\begin{equation}\label{eq:4times4finetuning}
 a^3_{12}\simeq-\left(\frac{\epsilon}{\Lambda}\right)^2a^1_{01}.
\end{equation}
Since the coefficients $a^1_{01}$ and $a^3_{12}$ can generically be either of
order unity or vanish exactly, it indeed follows from
Eq.~(\ref{eq:4times4finetuning}) that $T_1=T_3=0$, which is the criterion for
see-saw-Dirac particles in the case of $4\times 4$ matrices. Similarly,
in Sec. \ref{sec:6x6}, one can
confirm the validity of the criterion for see-saw Dirac
particles in the case of $6\times 6$ matrices.

\section{Summary and conclusions}
\label{sec:sc}

In conclusion, we have discussed general properties of neutrino mass matrices
involving a fine-tuning condition and the connection to flavor
symmetries. We especially pointed out that the number of light
neutrino masses generated by the see-saw mechanism cannot exceed half
of the dimension of the considered mass matrix if fine-tuning is 
absent. Furthermore, we have introduced the concept of
see-saw-Dirac particles. In the light of this, we formulated for
the examples of real symmetric $4\times 4$ and $6\times 6$ neutrino mass
matrices a necessary and sufficient criterion in order
to obtain see-saw-Dirac particles. For these cases it was shown that the
imposition of a conserved ZKM charge is equivalent to the assumption that the
mass terms in flavor basis represent independent parameters. As an
application of our methods, we demonstrated that small pseudo-Dirac
neutrino masses can be generated in a natural way by the see-saw
mechanism if discrete or non-Abelian symmetries are taken into
account. Then, we presented a model scheme based on one continuous
non-Abelian symmetry and one discrete Abelian symmetry generically
leading to a $6\times 6$ mass matrix, which fulfills the criterion for
see-saw-Dirac particles in the absence of a conserved ZKM
charge. Furthermore, we have found that a 
considerably wide class of reducible representations of unbroken
unitary flavor symmetries accounts only for the degeneracy of some
neutrino masses, but does not establish any relations between the
non-degenerate neutrino masses. Finally, we confirmed the above
formulated fine-tuning condition, {\it i.e.}, the criterion for
see-saw-Dirac particles coming from our symmetry considerations.

\begin{acknowledgments}
We would like to thank E.~Kh.~Akhmedov, M.~Freund, and W.~Grimus for
useful discussions and valuable comments.
Support for this work was provided by the Swedish Foundation for International
Cooperation in Research and Higher Education (STINT) [T.O.], the
Wenner-Gren Foundations [T.O.], and the ``Sonderforschungsbereich 375
f{\"u}r Astro-Teilchenphysik der Deutschen Forschungsgemeinschaft''
[M.L., T.O., and G.S.].
\end{acknowledgments}

\bibliography{references_r}

\end{document}